# Evaluating Learning Games during their Conception

Iza Marfisi-Schottman[1], Sébastien George[1], Franck Tarpin-Bernard[2]

[1] LUNAM Université, Université du Maine, EA 4023, LIUM,
Avenue Olivier Messiaen, 72085 Le Mans, France

[2] Université Grenoble Alpes, LIG, F-38000 Grenoble,
CNRS, F-38000, Grenoble, France

iza.marfisi@univ-lemans.fr
sebastien.george@univ-lemans.fr
franck.tarpin@ujf-grenoble.fr

**Abstract:** Learning Games (LGs) are educational environments based on a playful approach to learning. Their use has proven to be promising in many domains, but is at present restricted by the time consuming and costly nature of the developing process. In this paper, we propose a set of quality indicators that can help the conception team to evaluate the quality of their LG during the designing process, and before it is developed. By doing so, the designers can identify and repair problems in the early phases of the conception and therefore reduce the alteration phases, that occur after testing the LG's prototype. These quality indicators have been validated by 6 LG experts that used them to assess the quality of 24 LGs in the process of being designed. They have also proven to be useful as design guidelines for novice LG designers.

**Keywords:** Learning Games, Serious Games, evaluation, quality, cost, design, conception, development

## 1. Introduction

Learning Games (LGs) are a sub-category of Serious Games that use game mechanics for educational purposes (Abt, 1970). Even though they have proven to be promising in many domains, where classical education technics fail (Dondlinger, 2007 ; Mayo, 2007), their use is at present restricted by their time consuming and expensive developing process. Indeed, it has been estimated that their cost usually varies between 10 and 300 thousand dollars (Aldrich, 2009) but some LGs, such as *America's Army* can cost up to 30 million dollars! In addition, LGs often focus on specific competencies that concern a very limited public. It is therefore very difficult to expect a positive return on investment. Effective LGs also need to be designed with a delicate symbiotic relationship between the educational and fun aspects, which is far from been easy to reach (Lepper and Malone, 1987).

In order to reduce the time and cost of LG design, we recommend a design methodology, inspired by industrial engineering theories, and based on frequent quality controls (Marfisi-Schottman et al., 2009). These controls allow comparing the LG, in the process of being designed, to the initial specification contract, in order to determine if it conforms to the client's needs. Such quality controls can help designers identify and repair eventual problems in the early phases of conception and therefore reduce the alteration phases that occur after testing the LG prototype.

In this article, we propose a set of quality indicators that can help designers evaluate the quality of their LG, during the designing process, and before it is developed. These indicators can also be given as guidelines to the designers in the early stages of conception. Even though these quality indicators are primarily aimed at quest-like LGs for teaching engineering skills, we believe the vast majority of them, found in literature, can be generalized to all types of LGs. In the second section of this article, we propose an evaluation protocol of these quality indicators. It involved 6 LG experts, who used them in order to evaluate the quality of 24 partially designed LGs. The quality indicators were also used and assessed by 12 novice LG designer, who used them as design guidelines.

## 2. Quality Indicators

In order to evaluate the quality of a LG, we propose a set of indicators, inspired by Lepper & Malone (1987) and Sanchez (2011) research but also by our lab's 20 years of experience in designing, developing and using LGs. In particular, we adapted the indicators found in literature to LGs used in higher education and to the fact that the LGs are still in the process of being designed and therefore cannot yet be tested by students. We also attempted to standardize the level of detail of these indicators in order to make them easily quantifiable by designers.

In order to analyze LGs with complementary angles, these indicators provide information concerning their educational and entertaining potential as well as their usefulness in educational contexts. They also allow measuring how close a LG is to the initial specification contract, and especially the fact that it covers all the given pedagogical objectives. Finally, they provide indications on the difficulties that might occur during the development phase. As show in figure 1, these indicators are structured according to 6 facets that represent complementary views of LG characteristics.

These facets are inspired by Marne and al.'s (2012) research. Indeed, after studying various analysis structures for LGs (Dempsey and Johnson, 1998; Ferdig, 2008), we chose theirs because we believed they offer the best solution to structure our indicators in categories that are easily understandable by designers. Nevertheless, we chose to merge the original "Domain Simulation" facet with the "Problems and Progression" facet because it seemed quite difficult to analyze these separately. We also added an extra "Estimated Cost" facet because reducing the production cost is one of your main concerns. In the next section, we will describe the 6 facets and their quality indicators.

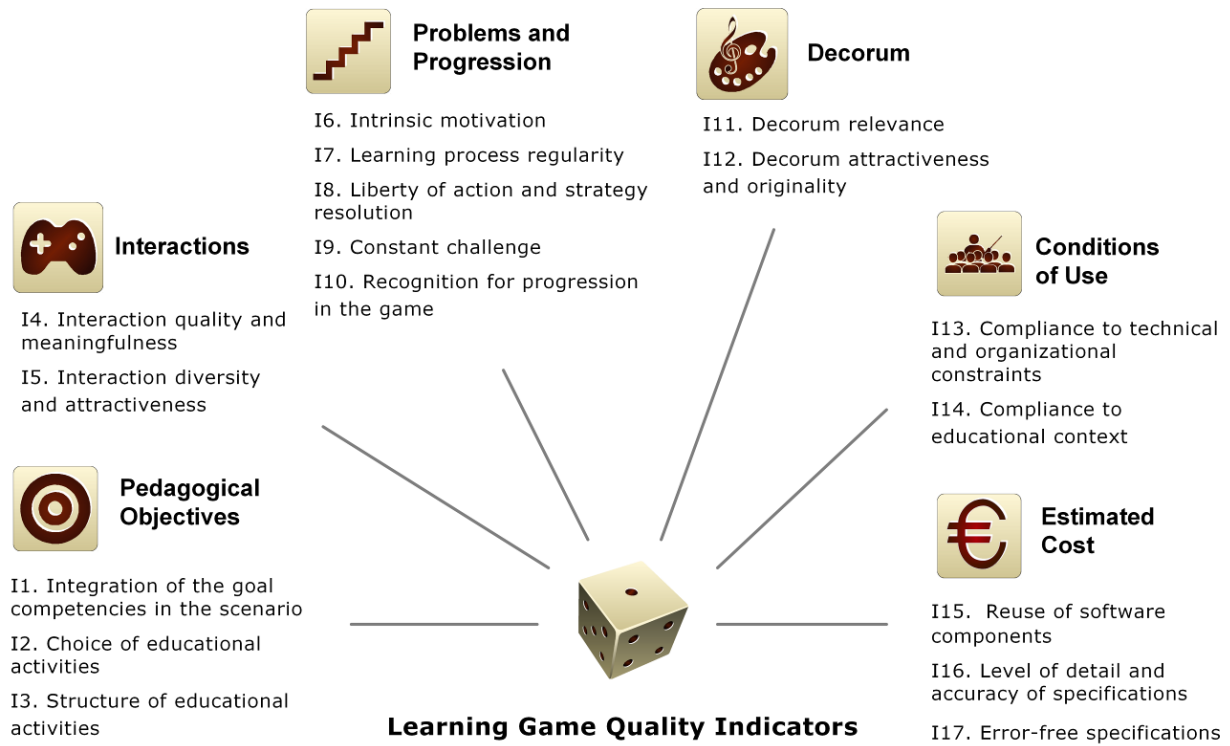

*Figure 1 • Quality Indicators structured in 6 Facets*

## 2.1. Pedagogical Objectives

This facet represents what the students need to learn (Marne and al., 2012). In order to analyze the LG's quality according to this facet, we propose several quality indicators that can provide hints on the LG's educational potential (the entertaining aspects will be analyzed in other facets).

### *I1. Integration of the goal competencies in the scenario*

In order for a LG to meet the client's pedagogical objectives, its scenario must integrate all of the identified goal competencies. Even if this condition is not sufficient to guarantee pedagogical effectiveness, we believe it is necessary to assure that all the goal competencies are addressed at least once.

### *I2. Choice of educational activities*

Competencies are acquired through various activities. In addition to activities such as reading and listening, found in traditional teaching settings, LGs offer more attractive activities in which learners have a more active role such as investigating, simulating, diagnosing, manipulating tools or creating. These activities need to be chosen wisely to fit the given educational situation (Dascălu et al., 2012). In order to analyze LGs from all angles, this indicator only focused on the relevance of this choice; the order in which the activities are structures is analyzed separately. There is no strict rule for this analysis, only specialists can determine the solutions that are the best adapted to a specific context, students and educational goals.

### *I3.     Structure of educational activities*

As mentioned above, the way the educational activities are structured and sequenced together needs to be meticulously planned according to the global pedagogical strategy (Dascălu et al., 2012). Is it best to teach the theoretical knowledge first or let the leaners test activities they will be unable to solve, in order to encourage them to lean by trial and error? Should the learners be evaluated at the beginning of the course, at the end or regularly, in order to engage and encourage them? Once again, there are no strict rule for this analysis and this is why we will let a specialist evaluate it.

## 2.2. Interactions

This facets represents all the interactions that learners will have with other human actors (learners, teachers) and computers devices (computer, phone, tablet…) during the LG (Marne et al., 2012). In the next part, we will describe the quality indicators linked to this facet.

### *I4.     Interaction quality and meaningfulness*

The quality of interactions can greatly enhance learner's motivation (Shneiderman, 1993). In addition, the choice of intuitive and meaningful Human-Computer interactions and mediated Human-Human interactions can facilitate LG's acceptance by teachers (Kirriemuir & Mcfarlane, 2004). The use of tangible objects is for example well adapted for teaching mechanical skills because they allow practicing specific movements with the real tools (e.g. screw driver, drill).

### *I5.     Interaction diversity and attractiveness*

In addition to the activities' quality and meaningfulness, it is also recommended to integrate several types of interactions in order to promote active pedagogy. In addition, the innovative aspect of technology can also have a positive impact on learner's motivation (Daniel et al., 2009). This is the case of many recent LGs featuring smartphones, geolocation and augmented reality (Huizenga et al., 2007).

## 2.3. Problems and Progression

This facet represents the problems with which the learners will be confronted and the progression mechanics that will lead them to the next game level (Marne et al., 2012). In the next part, we will describe the quality indicators that can provide information on the meaningfulness of these problems and progression mechanics.

### *I6.     Intrinsic motivation*

It has been shown that LG's educational outcome is enhanced by intrinsic motivation (Habgood, 2007).

This type of motivation implies that the learning process is exclusively driven by the interest and the pleasure that the learners find in the LG, without the use of external rewards or punishments. In order to create this intrinsic motivation, the pedagogical activities must be meticulously woven to the game scenario. In other words, the environment, the interactions and progression mechanics in the game need to be closely related to the educational activities and goals (Habgood, 2007; Ryan and Deci, 2000).

### *I7.     Learning process regularity*
In the context of class-room training, teachers rarely seem to think they have too much time. We therefore believe that LGs used in this context should not have long phases, unrelated to educational activities. However, this does not forbid to have short moments of pure entertainment in order to relax learners and increase their self-confidence before starting harder exercises.

### *I8.     Liberty of action and strategy resolution*
First of all, it is important for players to feel their actions in the game will not have an impact on real life (Brougère, 2005 ; Ryan & Deci, 2000). When the LG is used in the context of a course, for which students will be evaluated, this is not so easy to apply. However it is important to keep some moments in the LG where learners can play freely and learn by try-error. Secondly, giving learners the choice of the actions and strategies they use, is a good way to improve their autonomy and make them feel invested with a mission and responsibilities (Kirriemuir and Mcfarlane, 2004). This sense of liberty will hence increase their personal engagement in the LG (Habgood, 2007).

### *I9.     Constant challenge*
Challenge is a crucial element to captivate the learners' attention (Ryan and Deci, 2000). The LG's level of difficulty must therefore be neither to low, neither too high. In addition, according to (Csikszentmihalyi, 1990) theory of "flow", the level of complexity needs to increase progressively, all along the LG, so that learners always feel the need to reach higher objectives. One of the technics used so that players always feel capable of progressing is for example to give them several goals and subgoals, of various difficulty levels (Björk and Holopainen, 2004).

### *I10.    Recognition for progression in the game*
With LGs, even more so than in classical educational contexts, learners expect to be gratified when they succeed (Reeves, 2011). This recognition, expressed by scoring, trophies, congratulation messages, unblocking element of the games, gives them the sensation that they have done well so far and pushes them to continue their efforts. In addition, according to (Damasio, 1995) and the study lead by the (National Research Council, 2000), the emotional responses, triggered when players win, have a positive effect on their level of attention, their memory and also their capacity to make decisions.

## 2.4.  Decorum

This facet represents all the story and multimedia elements offered by LGs to entertain players (Marne et al., 2012). In the next part, we will therefore describe the quality indicators that can give us clues on whether the decorum has been chosen correctly according to the learners profile and the LG's context of use.

### I11. Decorum relevance

It is preferable to choose game settings, characters and missions for which the LG's educational goals are meaningful. This correlation favors knowledge and skill transfer to real situations (Ryan and Deci, 2000) and the acceptance of the LG by educators (Kirriemuir & Mcfarlane, 2004). This does not however rule out the choice of original decorum, that would suit the players better, rather than realistic simulations (Reeves, 2011).

### I12. Decorum attractiveness and originality

The virtual environment in which the game takes place needs to captivate the learners' attention. The games' attractiveness can be improved with visual and sound effect, humor or simply an original story line or graphics. It is also encouraged to introduce elements of surprise such as visual effects or an unexpected twist in the scenario, in order to keep learners in an active state and stimulate their emotions (Lazzaro, 2004). In addition, in order for leaners to feel concerned by the outcome of the game's story, it needs to correspond to their emotional profile (Lepper and Malone, 1987). When designing a LG for young adults for example, it is important to choose an environment and missions that will not be rejected by a part of these learners.

## 2.5.  Conditions of use

This facet represents the context in with the LG will be used. These context can range from a large variety of situation in class, at home, assisted by teachers, all alone or even in groups (Marne et al., 2012). In our context, the conditions of use are determined by the client, at the beginning of the project. In the next part, we will therefore describe the quality indicators that can provide clues on whether the designers' choices are adapted to these given conditions or not.

### I13. Compliance to technical and organizational constraints

In order for the LG to be usable, it's activities need to be compatible with the technical and organizational constraints such as available material (computers, tablets), the length of a course, the number of students and teachers that will use the game.

### I14. Compliance to educational context

In order for learners to appropriate themselves with the knowledge and skills seen during LGs, and understand how they are relevant to their global learning curriculum, it is crucial for the LG sessions to be clearly identified by the teachers (Djaouti, 2011). In addition, it is important to add debriefing sessions during which teachers reflect on the LG activities, identify the skills that have been constructed and discuss real situations in which they can be used again (Hadgood, 2007).

## 2.6. Estimated Cost

This facet represents the estimated cost, in terms of time and money, for the LG's design, development and also for setting it up and using it. In the next part, we will describe the quality indicators that can provide us with an idea of this expected cost.

### *I15. Reuse of software components*

The development cost can greatly be minimized by taking advantage of reusable software components. Depending on the resources accumulated by the designers, they could greatly benefit from using parts of their previous LGs or applications that have proven to be efficient, such as toolbars, multi choice questions or mini-games.

### *I16. Level of detail and accuracy of specifications*

The development time greatly depends on the level of detail and accuracy of the specification documents written by the designers. Indeed, if these documents are not clear or detailed enough, the development team will need to ask the designers for more details and further explanations and therefore loose precious time. In order to avoid this, the specification documents need to contain all the details necessary for the development team (i.e. developers, graphic designers, actor) to create the screens and media. In other words, the specification documents need to contain a description of each screen, with images or sketches, clear explanation of the user interactions (clickable objects) and the actions they will trigger, the sequence in which the screens appear, the characters and locations, described with an image or a sketch, and their dialogues. The specification documents also need to provide information to help teachers understand the LG and integrate it to their course in addition to all the necessary information to help the client justify the use of the LG for their educational purpose. All this information also obviously needs to be clear and well organized.

### *I17. Error-free specifications*

The game specifications must not contain any errors or misconceptions (e.g. unconnected screens, different names for the same character) that could block the development process and therefore increase the LG's development time.

In this precious section, we have proposed a set of 17 quality indicators that can be used to evaluate the potential quality of LGs during their design process. Once the LGs are developed, further evaluations can be conducted to assess their acceptability, usability and usefulness (Sanchez, 2011). For further tests, we also worked on simulators with virtual players that react according to predefined profiles (curious, cautious, fast…) (Manin et al., 2006). For the time being, these methods only exist for board games that have a very formal structure, but the principal should be extendable to other types of games. The objective of these simulations is to experiment the learner's progress through the LG, in order to statistically evaluate if the pedagogical objectives are attained. Now that we have described the quality indicators, we will clarify the context in which they have been used and validated by LG experts.

## 3. Quality Indicator Validation

The set of quality indicators presented in this article were evaluated in several ways: first they were given to 12 novice LG designers, as guidelines to help them design two LGs each (total of 24 LGs) and then, they we were used by 6 LG expert, to evaluate the potential quality of these 24 LGs. These evaluations actually took place in a wider context: measuring the utility of LEGADEE[1] (*LEarning GAme DEsign Environment*), our LG authoring environment (Marfisi-Schottman et al., 2010). Our goal was to verify that LEGADEE promoted the design of better quality LGs, more rapidly and with lower production costs. The results were promising, especially for several of the indicators but we will not give more details here, as this is not the central topic of this article. In the next part, we will discuss the two evaluations of the proposed quality indicators.

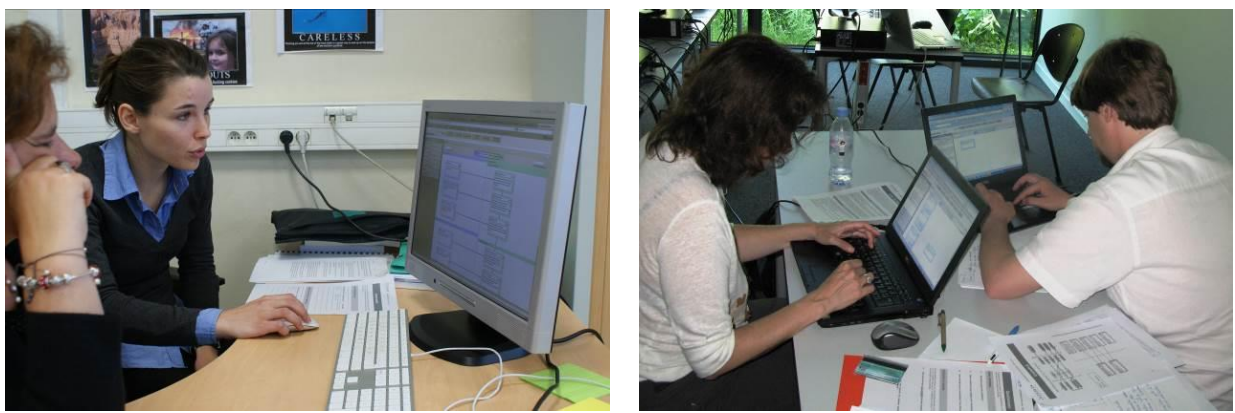

*Figure 2 • Pictures taken during the Learning Games design*

### 3.1. Use of quality indicators as guidelines

12 university teachers, in computer science, each designed two LGs for two specifications contracts (noted A and B) (Figure 2). LGs of type A need to meet the expectations of a university that wants a LG to help first year students to understand the basics of C programming. LGs of type B need to meet the expectations of a multinational food distribution company who needs a LG to teach their new recruits how to deliver groceries with electric scooters. In order to help these novice designers imagine their LGs, we provided them with the set of the quality indicators, under the shape of a "good practice" list with game examples. The design of each LG took between 1h10 and 4h20. In order to get their feedback, we used semi-conducted interviews during the design process and asked them to fill in a questionnaire at the end. In particular, to the question "**Did you read the good practice document? Did you find it useful and for what purpose?**" all the designers answered yes. Some of them also provided further explanations that show that this document was useful:

> *"I thought the good practice document was very helpful. I used it to verify I had done everything right before I handed in my game." – "The examples given in the good practices were a good way to get me started." – "I used it for inspiration."*

### 3.2. Use of quality indicators to evaluate LGs

The 24 LGs, designed by the teachers, where then evaluated by experts, with the quality indicators. In the context of our evaluation, and thanks to the data collected by the LEGADEE authoring environment, we were able to calculate the quantitative indicators (I1, I5, I15, I16 and I17) automatically. We asked 6 LG experts (researchers and professionals in this field) to evaluate the other 12 indicators. One expert had 5 years of experience in LG design, two had 8 years of experience and the others had 10, 13 and 20 years of experience. Each expert received an email with the description of the indicators, exactly as it they are given in the first part of this article, along with 8 LGs to evaluate (we distributed the LGs so that each one would be evaluated twice). They also received the two specification contracts A and B, so that they could judge if the LGs met the client's expectations. The experts had to evaluate each LG according to the 12 indicators, on a scale from 1 to 4. In addition, we added a specific zone next to each indicator, where they could add remarks, and we also asked them to fill out a questionnaire. The experts each worked between 1h45 and 3h in order to evaluate their 8 LGs.

The feedback provided by the experts underlined the fact that the LGs designed by the teachers where not precise and complete enough to conduct a thorough evaluation. This can be explained by the short design sessions (from 1 to 4h) that lead the teachers to stay at a very high level of design. The experts therefore had a hard time evaluating some of the indicators, as put forward by their answers to the question "**Did you have any difficulties to quantify some of the indicators?**":

> *"Yes, especially because the scenarios where sometimes very incomplete and we were tempted to interpret their meaning." – "Globally, I felt it was difficult to judge the games' quality at this stage of design." – "Sometimes difficult to evaluate, however, it was possible to proceed by comparison, by evaluating the scenarios according to one another."*

The experts also left comments directly on the evaluation grid that show the difficulty of evaluating certain aspects of the LGs:

---

[1] http://www-lium.univ-lemans.fr/legadee/

> *"Not very much info but I think I identified the principals of a race game"* – *"Very little information but seems OK"* – *"Even though it is not described, the integration of the LG to the education context seems meaningful."*

However, the experts' feedback on the quality indicators them-selves are very positive. Indeed, all the experts found that they were useful and easy to understand:

> *"The indicators are clear."* – *"The document on the indicators was very useful because it describes precisely what the needs are. Without it, the evaluation would have been much more difficult."*

Nevertheless, one expert expressed some reserves concerning the separation between the indicators *I1. Integration of the goal competencies in the scenario* and *I14. Compliance to educational context.* Another had a hard time understanding the indicator *E7. Learning process regularity*.

To the question « **Do you believe that other indicators could have given information on the potential quality of these LGs? If yes, with ones?** », four experts answered that they thought the grid was complete and sufficient. The two others suggested to add indicators concerning the diversity of game mechanics and the scenario's originality. Despite the fact that these characteristics are already in the indicators I4, I5 and I12, these comments show that they were not put forward or explicit enough.

Thanks to the experts comments, we were able to adjusted and improve the descriptions of several indicators in order to make them more clear and complete.

## 4. Conclusion and perspectives

In this article, we proposed a set of 17 quality indicators that help analyze the quality of LGs during their design process. The main idea behind this research is to facilitate quality controls throughout the design process, in order to minimize the time consuming and costly nature of LG's developing process. The evaluation of these indicators, involving six LG experts, allowed us to adjust and validate their terminology and validate the fact that they cover all the LG's important characteristics. Combined with game examples, these indicators also proved to be quite useful to guide and inspiring novices LG designers.

The use of these indicators during the evaluation process also brought to light the important amount of time that such an analysis can take. In order to make this evaluation process faster, we already calculated some of the quality indicators automatically, by analyzing the usage tracks provided by LEGADEE, the authoring environment in which the LGs where designed. Considering the positive results, we believe that this path could be pursued even further. The idea is not to replace the expert's evaluations, because we believe that certain indicators require human expertise, but to provide them with quantifiable elements, in order to help them make their decision. We therefore believe such indicators could be directly integrated into LG authoring environments, such as LEGADEE, in order to provide data and visual pointers to help experts or the designers themselves to validate their LGs, during their design process.